\documentclass[3p, 11pt]{elsarticle}
\usepackage{graphicx,amscd,amsmath,amssymb,verbatim}
\usepackage{amsfonts,epsfig}
\usepackage{mathptmx}
\usepackage{setspace}
\usepackage{epsfig}
\usepackage{url}
\usepackage{multirow}
\usepackage{color}

\begin{document}

\journal{Elsevier}

\begin{frontmatter}

\title{Distinct aggregation patterns and fluid porous phase in a 2D model for colloids with competitive interactions}

\author{Jos\'e Rafael Bordin} 

\address{Campus Ca\c capava do Sul, Universidade Federal
do Pampa, Av. Pedro Anuncia\c c\~ao, 111, CEP 96570-000, 
Ca\c capava do Sul, RS, Brazil

{\normalsize{E-mail:josebordin@unipampa.edu.br}}}

\begin{abstract}

In this paper we explore the self-assembly patterns in a two dimensional colloidal system
using extensive Langevin Dynamics simulations.
The pair potential proposed to model the competitive interaction 
have a short range length scale between first neighbors and a 
second characteristic length scale between third neighbors.
We investigate how the temperature and colloidal density will affect the assembled morphologies.
The potential shows aggregate patterns similar to observed in previous works,
as clusters, stripes and porous phase.
Nevertheless, we observe at high densities and temperatures a porous 
mesophase with a high mobility, which we name fluid porous phase, while
at lower temperatures the porous structure is rigid. 
triangular packing was observed for the colloids and pores in both solid and fluid porous phases.
Our results show that the porous structure is well defined for a large range of
temperature and density, and that the fluid porous phase is a consequence
of the competitive interaction and the random forces from the Langevin Dynamics.

\end{abstract}

\end{frontmatter}

\setlength{\baselineskip}{0.7cm}

\section{Introduction}

\label{Section:Introduction} 

The study of chemical building blocks as amphiphilic molecules, 
block copolymers, colloids and nanoparticles have attracted the attention in soft matter 
physics in recent years due their properties of self-assembly~\cite{White02, Roh05, Klapp16}.
The variety of the length scales and geometry of the patterns arise 
from the different types of potential energies involved and from the shapes of the colloidal particles.
Self-assembled nanomaterials have applications in medicine, self-driven molecules, catalysis, photonic crystals, 
stable emulsions, biomolecules and self-healing materials~\cite{Cas89, Talapin10,Zhang15, Velev00, Velev00b, Yin01}.

Despite the fact that the competition between shape and interaction 
can influence the self-assembly of molecules and colloids~\cite{Zdenek13, Ye13, BoK17a, Toole17},
experimental~\cite{Loku16, Peters17} 
and simulational studies~\cite{Zhuang16, Perez13,Lindquist16, Lindquist17, 
Pineros17, Archer07, Schwanzer16, Candia06, Godfrin13} have
show that spherical colloids with tunable competitive interactions can be
used to generate distinct aggregate patterns.
The combination of a short attractive interaction and a long-range repulsion 
can describe the nucleation in these systems with competitive interactions~ \cite{Stradner04, Shukla08}.
The repulsion can be caused by a soft shell, as in the case of polymeric brushes~\cite{Lafitte14, Curk14, Nie16}, or
by electrostatic repulsion in charged colloids and molecules~\cite{QP01, CA10, Ong15, Campos17}, while the 
attraction is caused by van der Walls forces or solvent effects~\cite{Alamarza14}.
The patterns formation in two-dimensions (2D) has been extensively 
explored in the literature by theoretical and computational
works~\cite{Toledano09, Pkalski14, Alamarza16, 
Muratov02,  Mossa04, Wu06, Imperio06, Imperio08,  Archer08,
Nasi16, Ciach13, Mendonza09, Patta15, Patta17, Zhao12, Zhao13}. 
These patterns includes small clusters, lamellae phases, stripes and 
porous phases.

In this paper we explore the phase diagram of a two dimensional colloidal system
with a short range attraction and a long range repulsion (SALR). 
This potential is based in a well known ramp-like core-softened potential, extensively applied in
studies of systems with water-like anomalies~\cite{Ol06a,Ol06b}. Here, we
parametrize the equation to reproduce a SALR potential with
an attraction at a distance $\approx \sigma$, the diameter of the disks,
and a second length scale at a distance $\approx 3\sigma$, with a energy barrier
between these characteristic distances. These 
distances were chosen to reproduce stripes with thickness $\approx 3 \sigma$, as
observed in charged globular proteins and nanoparticles~\cite{Shukla08, Kowa11}.
Here, our continuum SALR system is studied using extensive Langevin Dynamics simulations.
In previous works~\cite{BoK16a, BoK16b, BB17b} we have observed that Brownian forces
affects the phase diagram of systems with pattern formation.
These forces leaded to distinct self-assembled patterns,
distinct dynamical properties and even to a reentrant fluid phase.
Therefore, we can expect that new phenomena arises due the Brownian effects 
in the SALR system with these characteristic distances.

Our results show that the system has
a rich variety of patters, with clusters, stripes and porous
phases, similar to observed in lattice models~\cite{Alamarza14, Pkalski14}. 
Yet, here we observe a curious fluid porous phase that was not observed. 
In this phase, the 
pores have a well defined structure but the mean square displacement
shows that the particles are diffusing. 
This indicates that the pores have a fixed size and also diffuses, 
in a similar way to bubbles moving in a fluid. 

Our paper is organized as follows. In the Section~\ref{Model} we introduce our model and 
the details about the simulation method. On Section~\ref{results} we show and discuss our
results, and the conclusions are shown in Section~\ref{conclusions}.

%%%%%%%%%%%%%%%%%%%%%%%%%%%%%%%
\section{The Model and the Simulation details}
\label{Model}
%%%%%%%%%%%%%%%%%%%%%%%%%%%%%%%

In this paper we compute all the quantities using standard LJ units~\cite{AllenTild}. Distance, density of particles, time
and temperature are given, respectively, by

\begin{equation}
\label{red1}
r^*\equiv \frac{r}{\sigma}\;,\quad \rho^{*}\equiv \rho \sigma^{3}\;, \quad 
\mbox{and}\quad t^* \equiv t\left(\frac{\epsilon}{m\sigma^2}\right)^{1/2}\;, \quad
\mbox{and}\quad 
T^{*}\equiv \frac{k_{B}T}{\epsilon}\; ,
\end{equation}

\noindent where $\sigma$, $\epsilon$ and $m$ 
are the distance, energy and mass parameters, respectively.
In this way, we will omit the symbol $^*$ to simplify the discussion.

\begin{figure}[ht]
\begin{center}
\includegraphics[width=6cm]{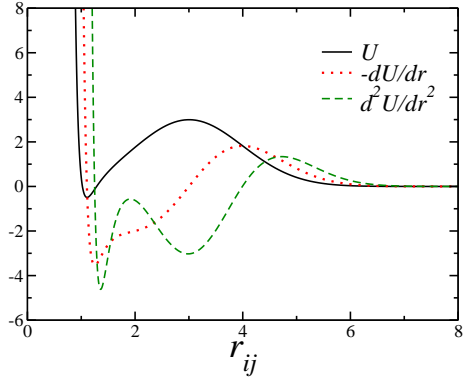}
\end{center}
\caption{Potential, force and second derivative of the potential
between two particles as function of their separation.}
\label{fig1}
\end{figure}

The colloidal system consists of $N = 2000$ disks with radius $\sigma$ and mass $m$ with a potential
interaction composed of a short-range attractive Lennard Jones potential and a Gaussian therm 
centered in $r_0$, with depth $u_0$ and width $c_0$, to take in account
the long-range repulsion
 $$
 U(r_{ij}) = 4\epsilon\left[ \left(\frac{\sigma}{r_{ij}}\right)^{12} -
 \left(\frac{\sigma}{r_{ij}}\right)^6 \right] + 
 $$
 \begin{equation}
 u_0 {\rm{exp}}\left[-\frac{1}{c_0^2}\left(\frac{r_{ij}-r_0}{\sigma}\right)^2\right]\;,
 \label{AlanEq}
 \end{equation}

\noindent where $r_{ij} = |\vec r_i - \vec r_j|$ is the distance between two colloids $i$ and $j$. 
This potential can be parametrized to have a ramp-like shape, and 
this particular shape was extensively applied to study systems with water-like properties~\cite{Ol06a, BoK15c}.
In this work, we propose the parameters $u_0 = 3.0$, $c_0^2 = 2$ and $r_0/\sigma = 3.0$.
The interaction potential is shown in figure~\ref{fig1}.
As showed by de Oliveira and co-workers~\cite{Charu10},the potential equation~(\ref{AlanEq}) have two length scales
associated with the minimum in their second derivative. In this way, the second derivative of
equation~(\ref{AlanEq}), dashed green line in the figure~\ref{fig1}, indicates that the potential
with the parameters proposed here have a first length scale close to $r_{ij}\approx1.0$ and a second
near to $r_{ij}\approx3.0$. Therefore, the pair interaction have
a short range first neighbors characteristic distance and a second length scale related to  
third neighbors. Usually colloids have interactions with a small range compared to 
their diameters, as extensively explored in the literature~\cite{Toledano09, Imperio08,Ciach13,Godfrin13}. 
However, in our model the range is much bigger than the disk size.
This is proposed to include the class of polymer-grafted colloids, where the 
polymer chain size and flexibility can lead to a long range in the interaction 
potential~\cite{Palli14, Wang16}.

Colloids are usually immersed in a solvent. Then, Brownian effects are relevant
for colloidal system, and an effective way to include solvent effects is by using 
Langevin Dynamics simulations.
In this way, Langevin Dynamics simulations were performed using the 
ESPResSo package~\cite{espresso1, espresso2}.
Hydrodynamics interactions were neglected.
Since the system is in equilibrium we do not expect that this will change the long-time
behavior.
The number density is defined as $\rho = N/A$, where $A= L^2$ is the area and $L$ the
size of the simulation box in the $x$- and $y$-directions.
$\rho$ was varied from $\rho = 0.05$ up to $\rho = 0.80$, and 
the size of the simulation box was obtained via $L = (N/\rho)^{1/2}$.
The cut off in the interaction, equation~(\ref{AlanEq}), 
is $r_{cut} = 8.0$. In all simulations, even for the higher densities,
$L > 2\times r_{cut}$.
The temperature was simulated in the interval between $T = 0.05$ and $T = 1.50$,
with viscosity $\gamma = 1.0$. 
The equations of motion for the fluid particles were integrated
using the velocity Verlet algorithm, with a time step $\delta t = 0.01$,
and periodic boundary conditions were applied in both directions.
We performed $1\times10^6$ steps to equilibrate the system. 
The equilibration time was then followed by $5\times10^6$ steps for the results 
production stage.
To ensure that the system was equilibrated, the pressure, kinetic 
and potential energy, number and size of aggregates 
were analyzed as function of time, as well several 
snapshots at distinct simulation times.
Once two dimensional systems are sensitive to the number of particles
in the simulation, we carried out simulations with 10000 colloids for some points,
and essentially the same results were obtained if $T \geqslant 0.10$. 
Also, we carried out
simulations with up to $5\times10^7$ steps, but the aggregates 
for most cases are stable and do not change after $\sim1\times10^5$ steps.
Nevertheless, even for the longer and bigger simulations, in some points at low temperatures, 
$T < 0.10$, and higher densities, $\rho > 0.55$, the energy was not well equilibrated.
These points were not used to construct the phase diagram.
Five independent runs, with distinct and random
initial positions and velocities for the colloids, were performed to ensure that
the patterns are not correlated to the initial configuration. 

The cluster size was analyzed based in the inter particle bonding~\cite{Toledano09}. Two colloids belong to the same cluster
if the distance between them is smaller than the cutoff 1.5. This value ensures that the force between the 
particles is near the minimum (as shown in the inset of the figure~\ref{fig1}).  The dependence of the cluster size
with the simulation time is then evaluated. With this, we obtain the mean number of colloids in each aggregate, $<n_c>$, 
in each simulation run.
If $<n_c>$ is smaller than the total number of colloids $N$, then the system is in a cluster phase.
However, if any cluster has spanned the simulation box in at least one of the directions, and if this did not change 
in time, we consider this as a percolating phase.
As well, if $<n_c>$ = $N$ then the system is in the percolating phase. 
Since this method is sensitive to the choice of the cutoff parameter, we also tested smaller values, 1.15 and 1.25,
and essentially the same result was obtained. To ensure the correctness of the method, for some cases we 
calculated manually $n_c$ from the snapshot to compare with the obtained from the cutoff parameter.
This showed that a higher value for the cutoff distance leaded to clusters bigger than 
the observed in the snapshots.

\begin{figure}[ht]
\begin{center}
\includegraphics[width=12cm]{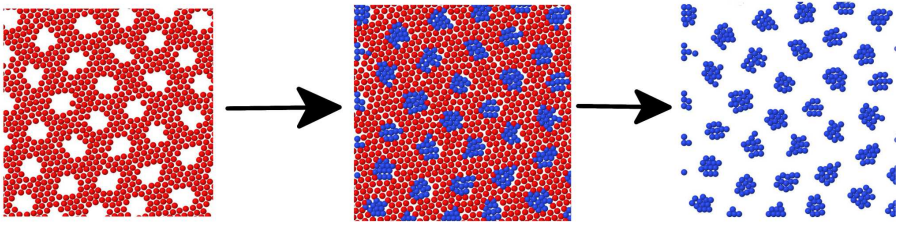}
\end{center}
\caption{Steps for the porous mesophase characterization. 
First, we select a snapshot of the system for a given temperature
and density. Here, the colloids are the red disks. Then, we attempt to insert ghost particles in a square lattice
with mesh 0.25. If there is no overlap a ghost (blue) disk is inserted. With the positions of the ghost particles we can
evaluate the porous properties.}
\label{fig2}
\end{figure}

To characterize the ``holes'' in the porous mesophase we take a collection of 1000 snapshots of each simulation run
and attempt to insert ghost particles with diameter $\sigma$ in a square lattice with mesh size 0.25.
If there is no overlap with the colloids or with ghost particles already inserted 
we insert a new ghost particle in the position. With the positions of the ghost particles, 
the same criteria used to characterize the colloidal aggregates is employed to characterize
a pore, but with a distance criteria equal to 1.0. 
We consider as a pore when the hole is filled with at least 4 ghost particles. Pores smaller than
this was considered as small defects and did not account for analysis.
With this, we can evaluate properties as the pore area $A_p$ and 
the radial distribution function $g_{p-p}$ between the center of mass of each pore.
In figure~\ref{fig2} we show a example of this construction.

To study the dynamical properties of the system we analyze the relation between 
the mean square displacement (MSD) and time, 
%%%%%%%%%%%%%%%%%%%%%%%%%%%%%%%%%%%
\begin{equation}
\label{r2}
\langle [\vec r(t) - \vec r(t_0)]^2 \rangle =\langle \Delta \vec r(t)^2 \rangle\;,
\end{equation}
%%%%%%%%%%%%%%%%%%%%%%%%
where $\vec r(t_0) = (x(t_0)^2 + y(t_0)^2)^{1/2} $ 
and  $\vec r(t) = (x(t)^2 + y(t)^2)^{1/2} $
denote the coordinate of a colloid
at a time $t_0$ and at a later time $t$, respectively.

%---------------------------------------------------------------------------------------------------
\section{Results and Discussion}
\label{results}
%---------------------------------------------------------------------------------------------------

\begin{figure}[t]
\begin{center}
\includegraphics[width=6cm]{fig3a.eps}
\includegraphics[width=6cm]{fig3b.eps}
\includegraphics[width=6cm]{fig3c.eps}
\includegraphics[width=6cm]{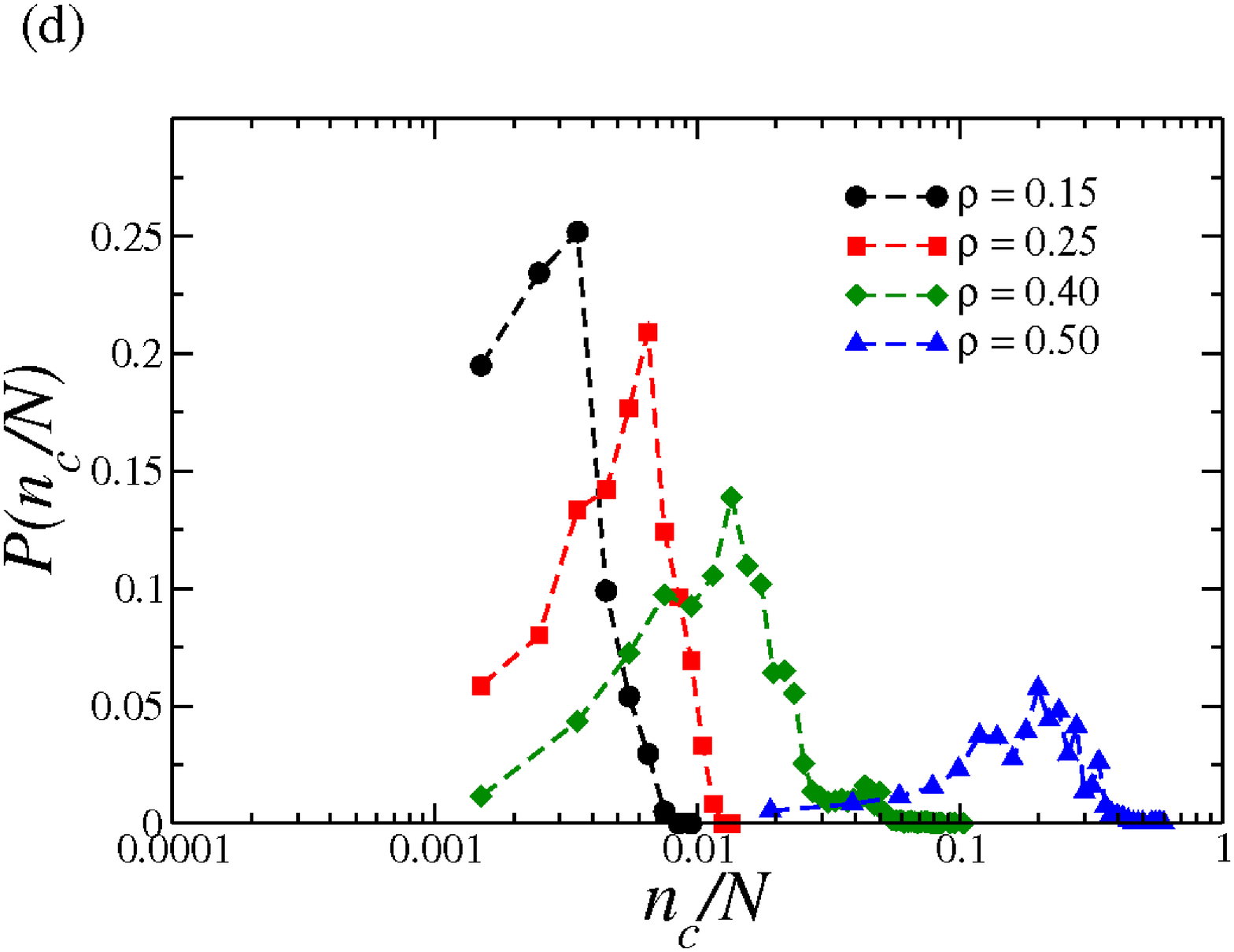}
\includegraphics[width=6cm]{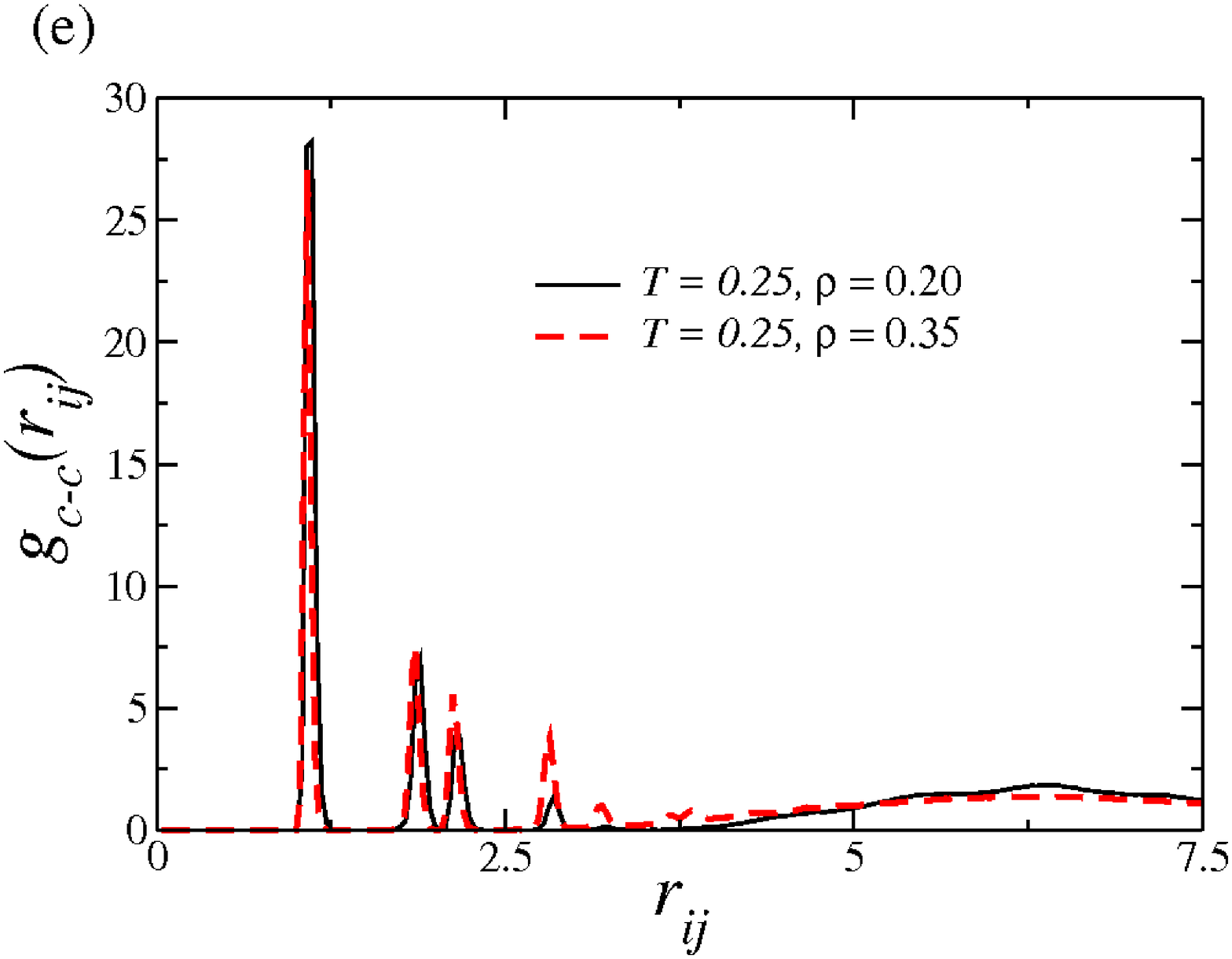}
\includegraphics[width=6cm]{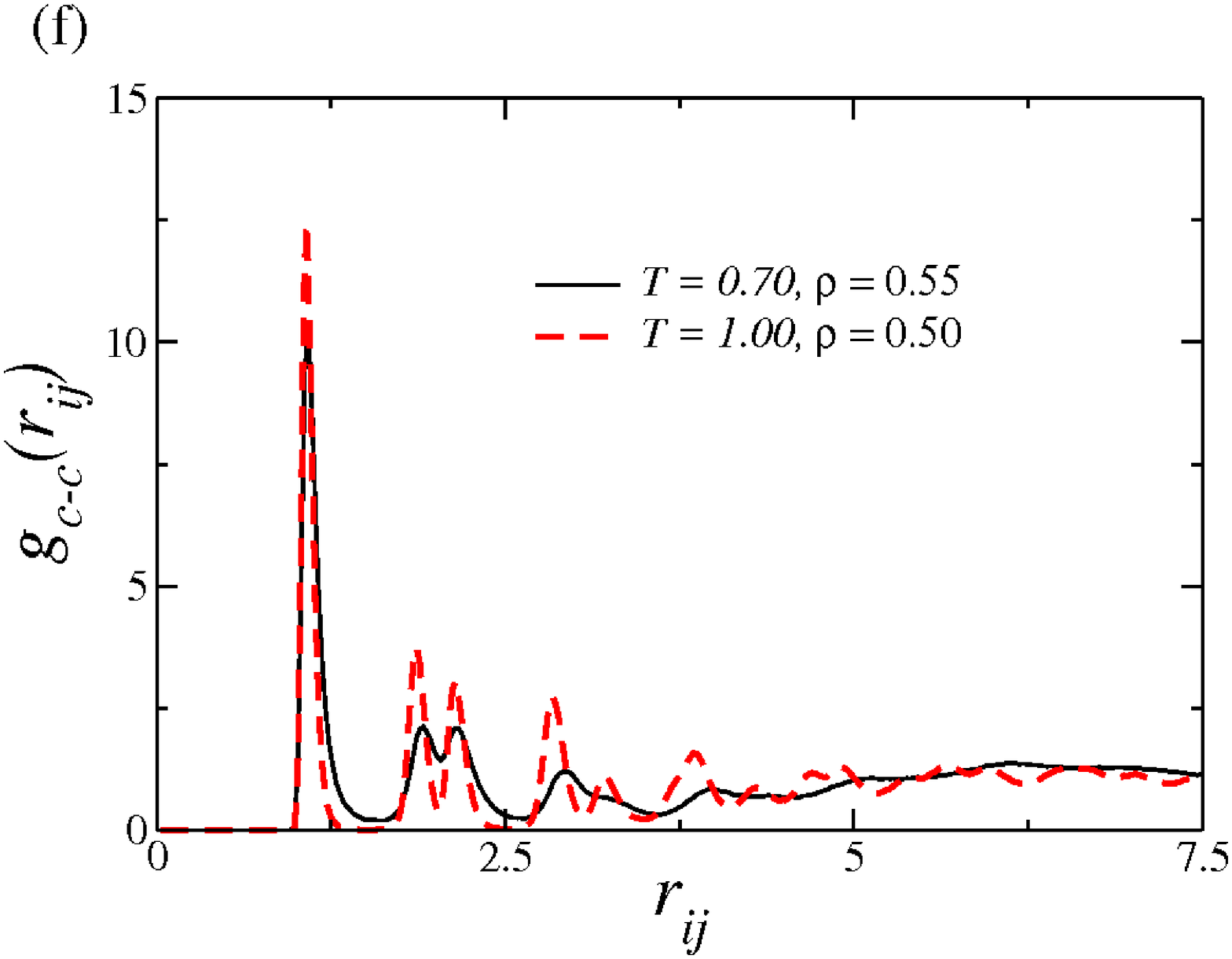}
\end{center}
\caption{(a) $T\times\rho$ phase diagram for the system. The gray dots are the simulated points. The dashed black lines 
divides the distinct aggregation regions. Region I is the cluster region, while regions II to VI are the percolating regions. These 
regions were divide accordingly with the observed structures and energy analysis. (b) Mean number of colloids in a aggregate as
function of the system density for distinct temperatures, showing the discontinuity when the system percolates. Inset: zoom in the
region at low densities. (c) Mean potential energy by particle, $u=U/N$, as function of the temperature for density $\rho = 0.75$. 
At this density the system is in the percolation region, but changes in the structure leads to gaps in the $u\times T$ curve.
(d) Probability $P(n_c/N)$ of find a cluster with size $n_c$ particles for the isotherm $T = 0.50$, showing the distribution
of cluster sizes in the non-percolating region. (e) Radial distribution function between colloids for two distinct
point inside the region I. (f) Radial distribution function between colloids for one
point inside the region II and one point inside region III.}
\label{fig3}
\end{figure}

In the figure~\ref{fig3}(a) we show the $T\times\rho$ phase diagram of the system. We should address that 
the phase diagram can be considered qualitative in a first moment, since the system can be metastable
in the coexistence lines. Therefore, the dashed lines indicates the separation between
regions with distinct patterns.
Nevertheless, as we will show, there are indicatives that these lines represents the separation
of the phases.The dashed lines divides the phase diagram in regions. First, there is the fluid region, where 
the colloids are not arranged in clusters and the MSD indicates that the system diffuses.
In this paper our focus is the aggregation patterns. In this way,
we divided the aggregate phase in six regions in the $T\times\rho$ phase diagram, as shown in the figure~\ref{fig3}(a). 
In each region a specific patterns was observed.

\begin{figure}[t]
\begin{center}
\includegraphics[width=12cm]{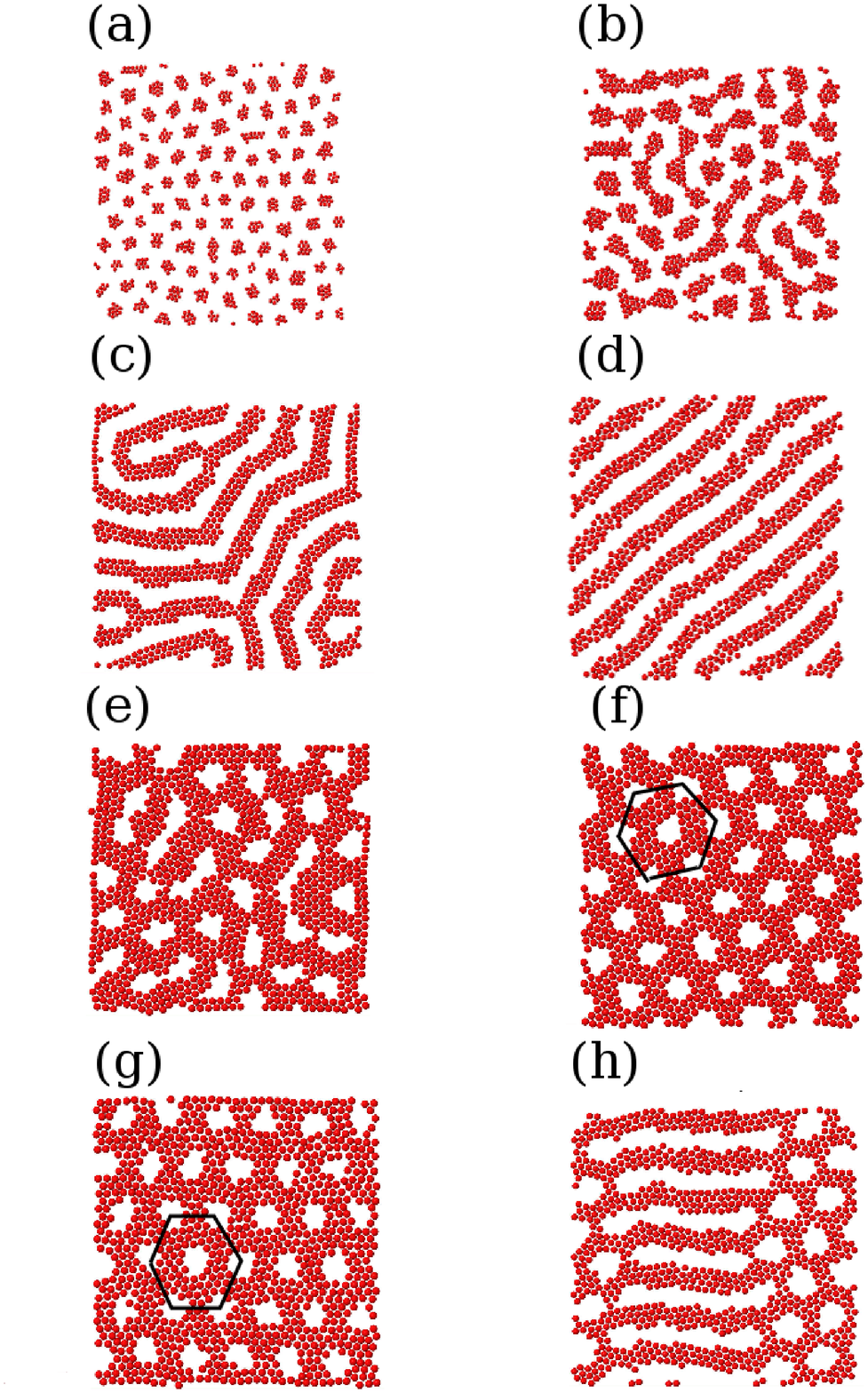}
\end{center}
\caption{Aggregate patterns observed in distinct regions. (a) Small clusters in region I, $T = 0.25$ and $\rho = 0.20$,
(b) big clusters in region I, $T = 0.25$ and $\rho = 0.34$, (c) disordered stripes in region II, $T = 0.70$ and $\rho = 0.55$,
(d) straight stripes in region III, $T = 1.00$ and $\rho = 0.50$, (e) porous mesophase with disordered pores in region
IV, $T = 0.10$ and $\rho = 0.75$,
(f) porous mesophase region V, 
$T = 1.00$ and $\rho = 0.75$, (g) porous liquid-crystal in region VI, $T = 1.10$ and $\rho = 0.65$,}
\label{fig4}
\end{figure}

The first step to define the pattern is calculate the mean cluster size.
In this way, the mean number of particles
in each cluster normalized by the total number of particles, $<n_c/N>$, was calculated 
and plotted as function of the density in figure~\ref{fig3}(b) for three isotherms. 
We have observed that
when $<n_c/N> < 1$ the colloids are aggregated in clusters, without percolation. 
The region I in the 
phase diagram restricts the points where we have observed these aggregates. The number of colloids in each cluster varies
accordingly with the temperature and density, as the inset in figure~\ref{fig3}(b) shows. 
Some examples of clusters in the region I are shown in the figures~\ref{fig4}(a) and (b).

Another indication of a transition from a clustering to a percolating phase in the 
plot of $<n_c/N>\times\rho$, figure~\ref{fig3}(b), is the discontinuity of the curves.
In the clustering phase, region I, $<n_c/N>$ increases linearly with $\rho$ at low densities.
Then, the curve have a discontinuity from the clustering densities ($<n_c/N> < 1$)
to the percolating densities ($<n_c/N> = 1$). This is a indicative of a phase transition from the clustering
to the percolating regions~\cite{BoK15c}. Then, the points $<n_c/N> = 1.0$ were considered as percolations
phases, and when $<n_c/N> < 1.0$ the colloids were in the cluster phase, region I. As we discuss later,
even a coexistence was observed in the points over the separation lines.
The cluster size distribution in the region I, that can be analyzed calculating the probability $P(n_c/N)$ 
of find a cluster with size $n_c$ particles, is shown in the figure~\ref{fig3}(d) for some isochores and the isotherm $T = 0.50$. 
As we can see, the probability of find a large cluster, and the mean size of the cluster,
increases with $\rho$. The $P(n_c/N)$ for the percolating region is not shown since all the particles
are in one single cluster in this region.

Inside the percolation regions distinct patterns can be observed.
To define the distinct patterns regions we have analyzed the system snapshots.
Since the mean size of the clusters can not be used to differentiate the distinct
patterns in the percolating phase, 
we have evaluated the mean potential energy by particle, $u=U/N$, 
and plotted it as function of the temperature. 
As an example, we show in the figure~\ref{fig3}(c) the curve of $u\times T$
the isochore $\rho=0.75$, that cross the regions IV, V and VI of the phase diagram, figure~\ref{fig3}(a).
As we can see, there are and slope changes and discontinuities at certain points. 
These gaps in the energy are related to changes in the aggregate
patterns.

Also, we can see structural changes analyzing the radial distribution function between 
the colloids, $g_{c-c}(r_{ij})$. 
Inside the region I, figure~\ref{fig3}(e), the cluster phase, the increase of the density
leads to a increase in the peak of the $g_{c-c}(r_{ij})$ near $r_{ij} \approx 3.0$,
while the figure~\ref{fig3}(f) shows that when the system goes from the percolating region II to percolating region III
the $g_{c-c}(r_{ij})$ indicates a higher ordering in the region III. This ordering becomes more clear
with the analysis of the system snapshots.
The first percolating structure is a disordered stripe pattern, shown in the figure~\ref{fig4}(c)
and observed in the region II of the phase diagram, figure~\ref{fig3}(a). The disordered stripe phase occurs
at lower temperatures than the aligned stripe morphology, shown in figure~\ref{fig4}(d), 
and enclosed by the region III in the phase diagram figure~\ref{fig3}(a).
The inter particle distance inside the stripe is equal to the disk diameter - the first length scale - 
and the distance between the stripes is three times the disk diameter, or the second length scale.
Increasing the density, the colloids rearrange from the stripe morphology to a porous phase. The shape of the pores 
are temperature dependent. For lower temperatures, the particles do not have kinetic energy to overcome the 
repulsive part of the potential to achieve the minimum in the energy. As consequence, the pores do not have circular 
symmetry and are randomly distributed. However, as usual for systems with two length scales~\cite{Ol06b, BB17b}, a increment in
the temperature can lead to a more ordered state. As consequence, at higher temperatures the pores have a circular
shape. The region IV in the phase diagram corresponds to the amorphous pore phase, while the
circular pore phase is inside the regions V and VI. A snapshot of these two distinct porous patterns are shown in the 
figures~\ref{fig4}(e), (f) and (g), for regions IV, V and VI, respectively. 

The sequence of patterns is similar to observed in previous works, as in the lattice model by Almarza and co-workers~\cite{Alamarza14},
Monte Carlo simulations by Imperio and Reatto~\cite{Imperio06} and theoretical approach by Archer~\cite{Archer08}. 
An interesting difference arises when compared to the lattice model, ref.~\cite{Alamarza14}.
In their work, the disordered lamellae pattern is observed at higher densities than the oriented lamellae pattern.
This is the opposite of the observed in our simulations. The difference came from the range of the potential. While in the 
work by Almarza {\it{et al.}} the second length scale is between second neighbors, in our case the second characteristic 
distance corresponds to the interaction between third neighbors. Then, our stripes are thicker, with three particles
side by side, while in the lattice model the stripes have a thickness of two particle. As consequence, 
the stripes at low densities are disordered, since is necessary a higher density to increase the 
stripe thickness and make it straight.

\begin{figure}[t]
\begin{center}
\includegraphics[width=13cm]{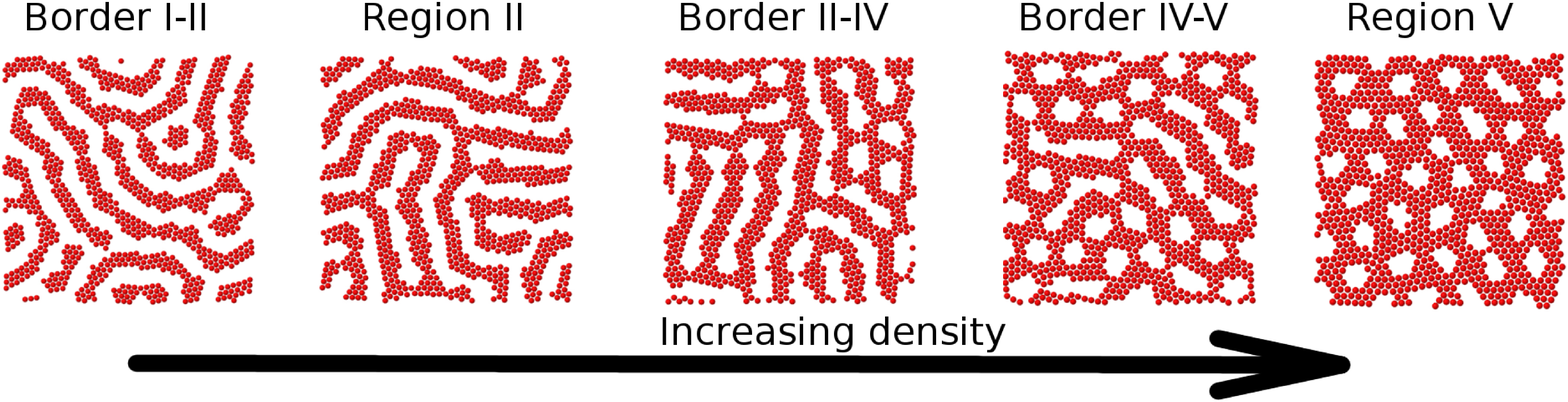}
\end{center}
\caption{Sequence of patterns for the isotherm $T = 0.50$ and densities $\rho$ = 0.50, 
$\rho$ = 0.55,$\rho$ = 0.60, $\rho$ = 0.65, and $\rho$ = 0.75 (left to right). The first snapshot
is at the border of regions I and II, and have patterns from both regions. The third snapshot
show the coexistence patterns observed in regions II and IV, while the fourth snapshot
have a pattern that is a mixture of the regions IV and V. The second and fifth snapshots
are points inside the regions II and V, respectively.}
\label{borders}
\end{figure}

\begin{figure}[t]
\begin{center}
\includegraphics[width=6cm]{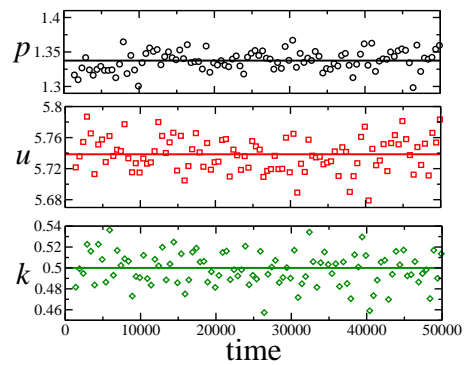}
\end{center}
\caption{Pressure $p$, energy per particle $u$ and the kinetic energy per particle
$k$ as function of time for the point $T = 0.50$, $\rho = 0.50$. Despite the point 
 being located in a coexistence region the thermodynamics quantities are well equilibrated
 and did not vary with time. The solid line represents the mean value.}
\label{time-dep}
\end{figure}

The lines that separate the regions have the discontinuity in the energy, 
the jump from $<n_c/N>< 1.0$ to $<n_c/N> =1.0$, and we can even observe
patterns coexisting in the snapshots. For instance, in the figure~\ref{borders} we show a sequence of patterns 
for the isotherm $T = 0.50$. As the plot in figure~ \ref{fig3}(b) indicates, at this isotherm 
the system percolates for $\rho > 0.55$. 
The first snapshot in figure~\ref{borders}, for $\rho = 0.50$, have two patterns
coexisting, large clusters and disordered stripes. 
Increasing the density to $\rho = 0.55$ only disordered stripes
were observed, in agreement with the transition from the cluster region I to the
percolating region II. Then, it is a indicative that at $T = 0.50$, $\rho = 0.50$ the system changes
from one pattern to another. Following the same isotherm, and increasing the density to $\rho$ = 0.60,
we can observe the coexistence of disordered stripes and amorphous porous, what we have identified as 
the border between the regions II and IV. In the same way, at $\rho$ = 0.65 there is a coexistence
of amorphous pores and circular pores. Finally, at $\rho$ = 0.75 all the pores have approximately 
the same diameter, characterizing the region V. Despite the apparent metastability in these points due the coexistence
of two patterns, we have indications that the system is well equilibrated.
For instance, we plot the pressure $p$, the energy per particle $u$ and the kinetic energy per particle
$k$ as function of time for the point $T = 0.50$, $\rho = 0.50$ in the figure~\ref{time-dep}. As
we can see, the energies and pressure did not vary with time, oscillating around the mean value.
As well, the patterns observed in the snapshots did not change with time.
Then, we can assume that the system is well equilibrated.

Next, comparing the patterns showed in the figures~\ref{fig4}(f) and (g) we should assume in a first moment that 
both are in the same region, since the two patterns correspond to a porous phase with circular pores.
Also, the pores are distributed in a triangular-like lattice in both cases, as the snapshots indicates.
Nevertheless, the energy analysis figure~\ref{fig3}(c) shows a discontinuity. 
Then, if the pattern is the same, what causes the energy jump?

\begin{figure}[t]
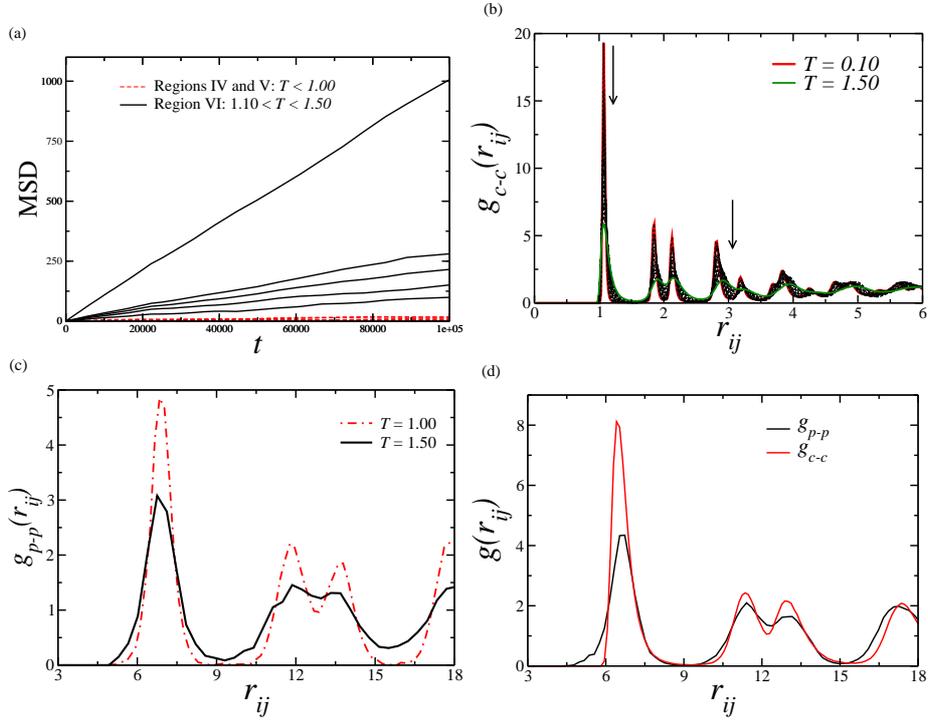

\begin{center}
\includegraphics[width=6cm]{fig5a.eps}
\includegraphics[width=6cm]{fig5b.eps}
\includegraphics[width=6cm]{fig5c.eps}
\includegraphics[width=6cm]{fig5d.eps}
\end{center}
\caption{(a) Mean square displacement as function of time for the density $\rho = 0.75$ and distinct temperatures, 
showing the transition from a solid to a liquid-crystal phase.  
(b) Radial distribution function for the colloids at $\rho = 0.75$ and for temperatures ranging from
$T = 0.10$ (red line) to $T = 1.50$ (green line).
(c) Radial distribution function 
between the pores center of mass at $\rho = 0.75$ and for densities $T = 1.00$ (region V) and $T = 1.50$ (region VI). 
(d) Comparison between the pores $g_{p-p}$ (black line) and the rescaled radial distribution function 
of the colloids $g_{c-c}$ (red line) at $T = 1.00$ and $\rho = 0.65$, showing that the pores packing follows the 
colloid packing.}
\label{fig5}
\end{figure}

To clarify this question, we have analyzed the colloids mean square displacement. The
analysis of the dynamical properties indicates that the colloids are diffusing above a temperature
threshold, as we show in figure~\ref{fig5}(a).
Therefore, the system is changing from a rigid porous structure to a fluid porous phase.
The fluid porous phase was designated as region VI in the phase diagram, figure~\ref{fig3}(a),
while the solid porous phase is the region V.

The Langevin thermostat employed in this study 
includes two therms in the 
net force a the disk $i$. The first is a drag force due the 
solvent viscosity, $-m\gamma \vec{v}_i$. The second is a
the random force $\vec{W}_i(t)$ due collisions between solute and solvent. 
This second therm is modeled as a Wiener process and
related to the system temperature $T$\cite{AllenTild}.
Therefore, as higher the temperature stronger is the random force. 
Is previous works for ramp-like core-softened fluids~\cite{BoK16b, BB17b},
we have shown that the Langevin thermostat leads to a melting and, as consequence,
to a reentrant fluid phase at high $T$.

Here, the entropic contribution from the random noise competes with 
the energy barrier in the potential interaction, figure~\ref{fig1}, to melt 
the system.
However, the energy barrier is high enough to ensure that the 
porous pattern will not be destroyed. Yet, we can see another changes in the 
colloidal system besides the increase in the MSD. For instance, analysis
of the radial distribution function between the colloids, $g_{c-c}(r_{ij})$, shown in
figure~\ref{fig5}(b), shows that for the solid ($T = 1.00$) and fluid ($T = 1.50$) porous phase  
the structure is similar, since the peaks are located at the same distances.
But looking the the valley between the peaks we can see that for the solid porous phase
this valley goes down, becoming zero. However, for temperatures in the fluid porous phase
the valley did not go to zero -- another indicative of a fluid porous phase.

Using the construction discussed in the Section~\ref{Model}, we can evaluate the center of mass (CM) 
from the imaginary ghost particles (here, all ghost particles have 
the same mass $m=1$, in LJ units) and evaluate the radial distribution function between the pores CM, 
$g_{p-p}(r_{ij})$. In the figure~\ref{fig5}(c) we show the $g_{p-p}(r_{ij})$ for samples in the regions V and VI.
We can see here the same behavior observed for the colloids: the peaks in the $g_{p-p}(r_{ij})$
are in the same distances for the solid and fluid porous phase. However, the valley goes to zero
only in the solid case. Then not only the colloids have the same pattern in regions V and VI, but the 
pores have the same distribution in both cases, regarding the presence or absence of diffusion.
To see the correlation between the colloid and pores radial distribution functions we rescaled the 
$g_{c-c}(r_{ij})$ curve. To do so, we consider that the first peak in the
$g_{p-p}(r_{ij})$ curve occurs at $r_{ij}\approx6.0$. Note that this is twice the thickness that we
have expected when the model was proposed, see the discussion in Section~\ref{Model},
and observed in the stripes of figure~\ref{fig4}(d). 
The rescaled colloid-colloid radial distribution function is compared with the pore-pore radial distribution function
in the figure~\ref{fig5}(d). As we can see, both curves have peaks at the same distances. So colloids and pores
have the same spatial distribution, with a triangular packing, as the snapshots in figure~\ref{fig4} indicates.
The same triangular packing is observed in the fluid porous phase.
Therefore, in region VI the system have a well defined porous structure (the peaks are well defined)
but have mobility (the valley did not go to zero) and diffuse (as the MSD curves show).

\begin{figure}[t]
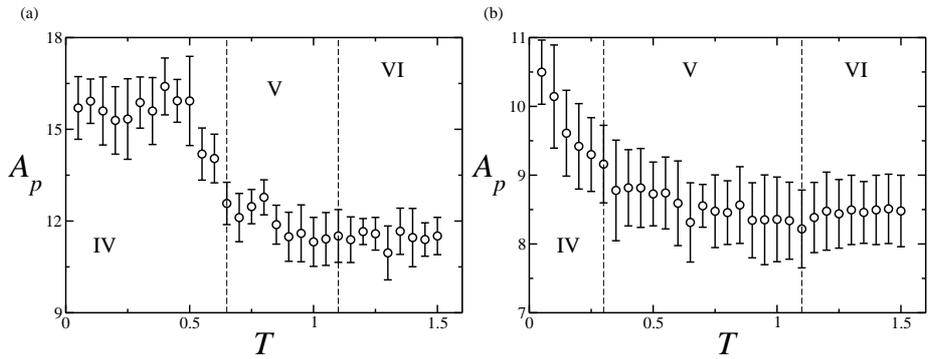

\begin{center}
\includegraphics[width=6cm]{fig6a.eps}
\includegraphics[width=6cm]{fig6b.eps}
\end{center}
\caption{Mean area of the pores as function of the the temperature for  
the isochores (a) $\rho = 0.65$ and (b) $\rho = 0.75$.  The dashed lines indicates the separation between the 
regions IV, V and VI in the $T\times \rho$ phase diagram.}
\label{fig6}
\end{figure}
  
Finally, we use the ghost particles to evaluate the mean area of each pore, $A_p$. This information give us a insight 
about the porous media properties to be used in technological applications, as filtration. As expected, at low 
temperatures the $A_p$ is larger once the system is in the region IV, with amorphous pores. In the solid region V
and the liquid-crystal region VI the pore area did not vary with the temperature. However, there is a small dependence
with the density. In figure~\ref{fig6} we show the curves for the densities $\rho = 0.65$ and $\rho = 0.75$.
For the lower density the area for the circular pores is $A_p\approx11.5$. Using the equation for the area of a circle,
$A = \pi a^2$, where $a$ is the radius of the circle, the pores have a mean radius of $a\approx1.91$.
At higher density $\rho = 0.75$, where $A_p\approx8.5$, $a\approx1.65$. Then, the difference in the pore radius
due the increase in the density is small. As well, once inside the circular porous region, the pores radii
did not vary with the temperature. Therefore, the 2D porous membrane is well structured, with a well defined
porous radius, inside a large range of densities and temperatures.

%---------------------------------------------------------------------------------------------------
\section{Conclusion}
\label{conclusions}
%---------------------------------------------------------------------------------------------------
To conclude, in this paper we have employed extensive Langevin Dynamics simulations to
explore the phase diagram of a colloidal system with competitive interactions.
The potential interaction has an attraction between first neighbors and a second length scale interaction
between third neighbors.
A large variety of patterns was observed. Despite the similarity with the structures observed 
in other works, the potential shape used in this paper associated with Brownian Dynamics effects
leads to a fluid porous phase. In this fluid porous phase the structure is well defined,
but the colloids have mobility and diffuses.
The observed porous mesophase is stable, with a well defined size for the pores.
Also, due the characteristics of the potential -- distance between the length scales and size of the energy
barrier -- the porous mesophase is well defined in a long range of densities and temperatures. These findings
shade light in the recent advances for colloids engineered to spontaneously assemble in a desired nanostructure.

%---------------------------------------------------------------------------------------------------
\section{Acknowledgments}
%---------------------------------------------------------------------------------------------------

We thank the Brazilian agency CNPq for the financial support.

\section*{References}
% \bibliographystyle{elsarticle-num}
% \bibliography{biblioteca}

\end{document}